\documentclass[preprint,aps,prc,showpacs,nofootinbib]{revtex4-1}
\usepackage{epsfig}
\usepackage{graphicx,amsmath}

\newcommand{\PANDA}{$\overline{\rm P}$ANDA }

\def\phi{\varphi}

\newcommand\ba{\begin{eqnarray}}
\newcommand\ea{\end{eqnarray}}
\newcommand\be{\begin{equation}}
\newcommand\ee{\end{equation}}

\begin{document}

\author{E.~Tomasi-Gustafsson}
\title{Proton Form Factors: phenomenology}
\email[E-mail: ]{etomasi@cea.fr}
\altaffiliation{Permanent address: \it CEA,IRFU,SPhN, Saclay, 91191 Gif-sur-Yvette Cedex, France}
\affiliation{\it Univ Paris-Sud, CNRS/IN2P3, Institut de Physique Nucl\'eaire, UMR 8608, 91405 Orsay, France}

\begin{abstract}

A general description of proton form factors is presented, in the whole kinematical region. The existing data and selected phenomenological models are briefly discussed. 
\end{abstract}

\maketitle
\section{Introduction}


The electromagnetic structure of any particle of spin $S$ is parametrized in terms of $2S+1$  form factors (FFs). Protons and neutrons are described by two form factors. A deuteron (spin one particle) is described by three form factors, charge, electric, quadrupole. The $\alpha$ particle, spin zero, has one form factor. FFs are analytical functions of one kinematical variable, $q^2$, which  parametrizes the internal distance inside the nucleon. 

The traditional way to measure electromagnetic hadron FFs is based on elastic electron proton scattering $e^-+p\to e^-+p$ and on the annihilation reactions $e^++e^-\leftrightarrow p+\bar p$, assuming that the interaction occurs through the exchange of one virtual photon of mass $q^2$. These reactions are related by the symmetries which hold for the electromagnetic and strong interactions. Crossing symmetry states that the same amplitudes describe the corresponding scattering and annihilation channels. These amplitudes are in general complex functions of two kinematical variables, for example, the linear polarization of the virtual photon $\epsilon$ and the momentum transfer squared, $t=q^2=-Q^2$, or the total energy $s$ and the angle of one of the emitted particles, but these variables act in different regions of the kinematical space. As an example, for annihilation reactions, the time-like (TL) region, $q^2$ is positive and for scattering reactions, the space-like (SL) region, $q^2<0$. Due to unitarity, in the SL region hadron FFs are real, whereas, in TL region, they are complex functions of $q^2$.
 
The amplitudes which parametrize the vertex $\gamma^*NN$ are called $F_{1,2}$ the Dirac and Pauli form factors. A linear combination is also used, the Sachs electromagnetic FFs, electric $G_{E}=F_1-\tau F_2 $ and magnetic $G_{M}=F_1+F_2.$, where $\tau=-q^2/(4M^2)$, $M^2$ is the nucleon mass.

In the space-like region, in non relativistic approaches (and also in relativistic approaches, but only in the Breit frame) the Sachs electromagnetic nucleon FFs are the Fourier transforms of the charge and magnetic distributions inside the proton. In TL region, the center of mass system (cm) is the most well suited to the description of annihilation reactions. Here FFs can be interpreted as the time evolution of the charge and magnetic distributions \cite{Ku11}.

\section{The space-like region}

The characteristic that makes so powerful the description of the nucleon structure in terms of FFs, is that FFs contain all the dynamics of the reaction and depend on one variable only, the momentum transfer squared. The kinematics can be factorized out in such way that the unpolarized $ep$ cross section can be factorized in a term which corresponds to the Mott cross section (for relativistic scattering on point-like particles) and a factor which contains FFs and depend on $q^2$. The transfer momentum gives the internal size $r$ at which the nucleon is tested by a projectile of momentum $p$, through the relation $rp$=1. 

The Rosenbluth method consists in measurements of the elastic differential cross section at different angles for a fixed value of $Q^2$ \cite{Ro50}. The linear dependence of the reduced cross section (after extracting kinematical factors) as a function of $\cot^2(\theta/2)$ allows to determine $G_E(Q^2)$ and $\tau G_M(Q^2)$ as the slope and the intercept.

From unpolarized cross section measurements the determination of $G_E$ and $G_M$ has been done up to $Q^2\simeq 8.8 $ GeV$^2$ and $G_M(Q^2)$ has been extracted up to $Q^2\simeq 31$ GeV$^2$ under the assumption that $G_E=0$, and it is often approximated, for practical purposes, according to a dipole form: $G_D(Q^2)=  (1+Q^2/0.71\mbox{~GeV}^2)^{-2}$.   

\begin{figure}
\begin{center}
\includegraphics[width=70mm]{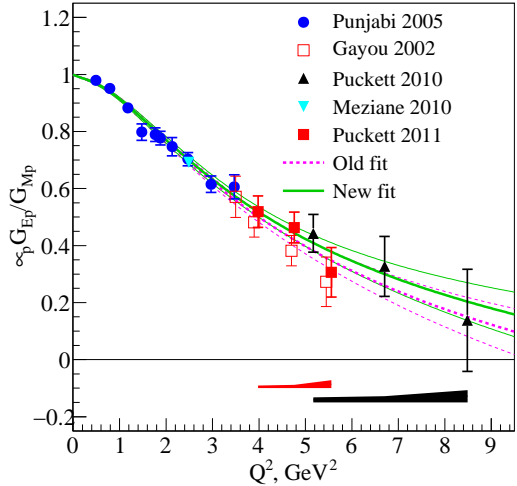}     
\caption{ Data on the proton FF ratio  $\mu G_E/G_M$ , as function of $Q^2$, from the recent polarization measurements (Ref. \protect\cite{Pu10}).}
\label{Fig:GEp}
\end{center}
\end{figure}

In recent years, very surprising results have been obtained, due to the possibility to apply the polarization method suggested in the sixties by A. I. Akhiezer and M. P. Rekalo \cite{Re68}. The GEp collaboration at JLab did not find a $q^2$ dependence of FFs compatible with a dipole form, but, instead,  a monotonic decrease of the ratio $\mu G_E/G_M$ with $q^2$, deviating from unity as $q^2$ increases, see Ref. \cite{Pu10} and refs therein (Fig. \ref{Fig:GEp}).


The difference of the ratio from unity is attributed to the electric FF, as the magnetic contribution is assumed to be well known from the cross section (at large momentum transfer it represents more than 90\%). The discrepancy between the data extracted from two different methods is likely to be attributed to radiative corrections. The probability to irradiate one or more photons from a few GeV electron (initial, and final) may reach 40$\%$. Radiative corrections are applied at first order to unpolarized data and are neglected in polarization experiments, as they factorize and cancel (at first order). Higher order corrections should be included. The lepton structure function method, proposed in Ref. \cite{Ku85} can be successfully applied in this domain \cite{By07}.

\section{The time-like region}

The annihilation  processes $ \bar p + p\leftrightarrow e^++ e^- $
allow to access the time-like region, over the kinematical threshold, $q^2>4M^2$. The differential cross section for $ \bar p + p\to e^++ e^-$, first derived in Ref. \cite{Zi62} in cm system, can be rewritten as a linear function of $\cos^2\theta$.

This results directly from the assumption of one-photon exchange, where the spin of the photon is equal to one and the electromagnetic hadron interaction satisfies $C$ invariance. Similarly to the the Rosenbluth fit, any deviation from linearity can be attributed to contributions beyond the Born approximation, as two photon exchange or photon radiation. 

The individual determination of the FFs in time-like region is possible through a precise measurement of the angular distribution (which is equivalent to the Rosenbluth separation in SL region). Due to the lack of statistics, mainly due to the luminosity achieved in colliders, only few data points on the FF ratio are available in the region above threshold.  The 
experimental results are usually given at a fixed value of $s=q^2$, 
in terms of a generalized FF, $|F_P|$ extracted from the angle integrated cross section, under the hypothesis that $G_E=0$ or $|G_E|=|G_M|$. The first hypothesis is 
arbitrary whereas the second one is strictly valid at threshold. Similarly to the SL region, $G_E$ plays a minor role in the cross section as $q^2$ increases. Therefore, different hypotheses for  
$|G_E|$ do not affect strongly the extracted values of $G_M$, due to 
the kinematical factor 
$\tau$, which weights the magnetic contribution to the differential cross 
section. 

Nevertheless, the ratio $R=G_E/G_M$ has been determined from a two parameter fit of the differential cross section, by PS170 at LEAR \cite{Ba94}, and more recently  by the BABAR Collaboration using initial state radiation (ISR), $e^++e^-\to \overline{p}+p+\gamma$ \cite{Babar}. If the emitted photon is sufficiently hard, one can factorize out from the cross section of this process, a factor which depends only on the photon variables. The results from BABAR suggest a ratio larger than one, in a wide region above threshold, whereas data from \cite{Ba94} suggest smaller values. Data are affected by large errors, mainly due to statistics.

At the future complex accelerator FAIR, in Darmstadt, the \PANDA collaboration \cite{Panda} plans to measure TL FFs through the annihilation reaction $ \bar p + p\to e^++ e^-$, 
using an antiproton beam of momentum up to 15 GeV and luminosity ${\cal L}=2\cdot 10^{32}$ cm$^{-2}$s$^{-1}$.
\PANDA is a $\sim 4\pi$ fixed target detector, designed to achieve momentum resolution at percent level for charged particles, high rate capability up to 10 MHz and good vertex resolution ($\sim$100 $\mu$m). The individual determination for FFs can be done up to  to $q^2\sim 14$ (GeV/c)$^2$, and, with a precise knowledge of the luminosity, the absolute cross section can be measured up to $q^2\sim 28$ (GeV/c)$^2$ allowing to extract the generalized FF, $|F_P| $. An improvement of at least one order of magnitude is expected, compared to the existing data \cite{Su10}.

Let us mention the interest of TL low $q^2$ region, near or under the kinematical threshold. Following an idea of M.P. Rekalo \cite{Re65}, one may reach the 'unphysical region' ($0<q^2<4M^2$) with three-body reactions, such as  $\bar p+p\to e^++e^-+\pi^0$ which is sensitive to nucleon electromagnetic and axial FFs \cite{Ad07}. 

The cross section for $e^++e^-\to p+\bar p(\Lambda+\bar\Lambda)$ does not vanish at threshold, but it shows a constant behavior in a wide region. This has been interpreted as an evidence that the hadron at threshold behaves as a point-like particle \cite{Ba10}. The near threshold region will be precisely investigated by the BES collaboration.

\section{Discussion and Conclusions}

As stressed in the Introduction, FFs are fundamental quantities: they are directly related to experimental observables, on one side, and on the other side, they parametrize the hadronic current. therefore any nucleon model, after reproducing static properties as masses and magnetic moments, should be tested on nucleon FFs.


\begin{figure}
\begin{center}
\includegraphics[width=75mm]{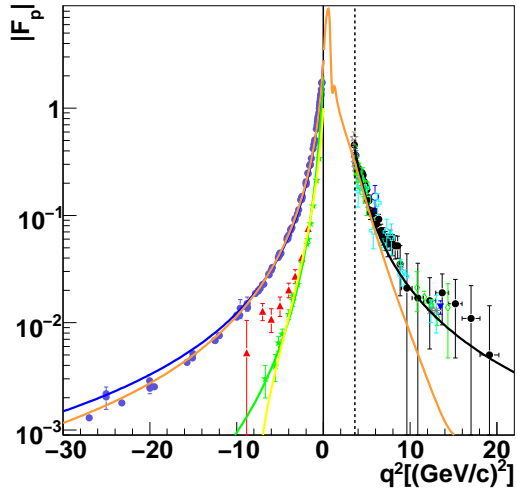}     
\caption{ World data on proton FFs as function of $q^2$ from Ref. \protect\cite{Ku11}. Space-like region: $G_M$ data (blue circles), dipole function (blue line); electric FF, $G_E$, from unpolarized measurements (red triangles) and  from polarization measurements (green stars). The green line is a monopole prediction for the ratio $G_E/G_M$. Time-like region ($q^2>4M_p^2$):  $|G_E|=|G_M|$ (various symbols). Shifted dipole (black line); prediction from VDM model \cite{Wa04} (yellow line).}
\label{Fig:All}
\end{center}
\end{figure}
Most nucleon models have been built to describe SL data and sometimes do not contain the analytical properties which are required to describe the TL region. Efforts for extending constituent quark models have been recently done. Models based on dispersion relations naturally describe the complex nature of FFs in TL region. Vector Dominance Models (VDM), which imply that the interaction occurs through the exchange of a vector meson of the same quantum numbers as the virtual photon ($\rho$, $\omega$, $\phi$) reproduce quite well the data, at the price of a number of parameters \cite{Wa04}. Measurements in the subthreshold region, which is expected to contain a huge contribution from the vector meson resonances would strongly constrain these models. 

In general, nucleon models are presently little constrained and the quantitative predictions display a large dispersion. The models originally built in the SL region which can be  analytically extended to the TL region may be readjusted to fit the world data in all the kinematical region ($i.e.$, in SL region, the electric and magnetic proton and neutron FFs, and in TL region, the magnetic FF of the proton and the few existing data for neutron).  Although these models may reproduce reasonably well the FFs world data, after a fitting procedure, they give different predictions in the kinematical regions where data are not available, and particularly for all polarization observables. 

Several experiments are planned or ongoing in electron accelerators as JLab, Mainz and colliders as Novosibirsk, BES, and Panda at FAIR. In SL region, the main purpose is to reach higher transferred momenta or better precisions. In TL region the individual determination of the electric and magnetic FFs has the highest priority. It is also foreseen the measurement of polarization observables, which would allow to determine the relative phase of FFs, which are complex functions in TL region.

\acknowledgments
M.P. Rekalo is at the origin of many ideas reported in this contribution. Thanks are due to E.A. Kuraev and G.I. Gakh for their collaboration, useful remarks and fruitful discussions. S. Pacetti is acknowledged for attentive reading and suggestions.


\begin{thebibliography}{0}
\bibitem{Ku11}
  Kuraev~E.~A., Tomasi-Gustafsson~E., Dbeyssi~A.,
  Phys.\ Lett.\ B {\bf 712} (2012) 240

 \bibitem{Re68}
  A.~I.~Akhiezer and M.~P.~Rekalo,
  Sov.\ Phys.\ Dokl.\  {\bf 13} (1968) 572
  [Dokl.\ Akad.\ Nauk Ser.\ Fiz.\  {\bf 180} (1968) 1081];  
  Sov.\ J.\ Part.\ Nucl.\  {\bf 4} (1974)  277
  [Fiz.\ Elem.\ Chast.\ Atom.\ Yadra {\bf 4} (1973) 662].
\bibitem{Pu10}
  A.~J.~R.~Puckett {\it et al.},
  Phys.\ Rev.\ Lett.\  {\bf 104} (2010) 242301 and Refs therein.

 
\bibitem{Ro50} 
M. N. Rosenbluth, 
 Phys. Rev. {\bf 9} (1950) 615. 


\bibitem{Ku85}
E.A. Kuraev and V. S. Fadin, 
Sov. J. Nucl. Phys. {\bf 41} (1985) 466
  [Yad. Fiz. {\bf 41} (1985) 733].

  \bibitem{By07}
 Yu. M. Bystritskiy, E. A. Kuraev and E. Tomasi-Gustafsson, 
  Phys. Rev. C {\bf 75} (2007) 015207.

\bibitem{Zi62}
A. Zichichi A., S. M. Berman, N. Cabibbo  and R. Gatto, 
 Nuovo Cim. {\bf 24} (1962) 170.

\bibitem{Ba94}
G. Bardin {\it et al.},
 Nucl.\ Phys.\  B {\bf 411} (1994) 3.

\bibitem{Babar}
B. Aubert {\it et al.}  [BABAR Collaboration], 
Phys.\ Rev.\  D {\bf 73} (2006) 012005.
    
 
\bibitem{Panda} 
The PANDA Collaboration, 
   preprint arXiv:0903.3905 [hep-ex]. 

\bibitem{Su10}
M. Sudol {\it et al.}, 
Eur.\ Phys.\ J.\  A {\bf 4} (2010) 373.

\bibitem{Re65} 
M. P. Rekalo, 
 Sov. J. Nucl. \ Phys.\ { \bf 1} (1965) 760.

\bibitem{Ad07}
C. Adamuscin C., E.A. Kuraev, E. Tomasi-Gustafsson and F.~E. Maas , 
Phys.\ Rev.\  C {\bf 75}(2007) 045205.
\bibitem{Ba10}
  R.~B.~Ferroli, S.~Pacetti and A.~Zallo,
  Eur.\ Phys.\ J.\ A {\bf 48} (2012) 33.


\bibitem{Wa04}
F. Iachello and  Q. Wan,
Phys.\ Rev.\ C {\bf 69} (2004) 055204.
  
\end{thebibliography}
\end{document}